\documentclass[floatfix,journal=nalefd,manuscript=letter]{achemso}
\setkeys{acs}{usetitle = true}
\usepackage{graphicx}
\usepackage{array}
\usepackage{amsmath,amsfonts,amssymb}
\usepackage[ansinew]{inputenc}

\newcommand{\Figref}[1]{Fig.~\ref{#1}}

\title{Pulling and Stretching a Molecular Wire to Tune its Conductance}

 \author{Gaël Reecht}
\affiliation{IPCMS de Strasbourg, UMR 7504 (CNRS -- Universit\'e de Strasbourg), 67034 Strasbourg, France}
\author{Hervé Bulou}
\affiliation{IPCMS de Strasbourg, UMR 7504 (CNRS -- Universit\'e de Strasbourg), 67034 Strasbourg, France}
\author{Fabrice Scheurer}
\affiliation{IPCMS de Strasbourg, UMR 7504 (CNRS -- Universit\'e de Strasbourg), 67034 Strasbourg, France}
\author{Virginie Speisser}
\affiliation{IPCMS de Strasbourg, UMR 7504 (CNRS -- Universit\'e de Strasbourg), 67034 Strasbourg, France}
\author{Fabrice Mathevet}
\affiliation{Institut Parisien de Chimie Moléculaire, Chimie des Polymères, UMR 8232, (CNRS -  Université Pierre et Marie Curie), 75252 Paris, France}
\author{C{\'e}sar Gonz{\'a}lez}
\affiliation{Service de Physique de l'Etat Condens\'e (CNRS URA2464), IRAMIS, CEA Saclay, 91191 Gif-Sur-Yvette, France.}
\author{Yannick J. Dappe}
\affiliation{Service de Physique de l'Etat Condens\'e (CNRS URA2464), IRAMIS, CEA Saclay, 91191 Gif-Sur-Yvette, France.}
 \author{Guillaume Schull}
\affiliation{IPCMS de Strasbourg, UMR 7504 (CNRS -- Universit\'e de Strasbourg), 67034 Strasbourg, France}
\email{schull@unistra.fr}

\begin{document}

\begin{abstract}
A scanning tunnelling microscope is used to pull a polythiophene wire from a Au(111) surface while measuring the current traversing the junction. 
Abrupt current increases measured during the lifting procedure are associated to the detachment of molecular sub-units, in apparent contradiction with the expected exponential decrease of the conductance with wire length. \textit{Ab initio} simulations reproduce the experimental data and demonstrate that this unexpected behavior is due to release of mechanical stress in the wire, paving the way to mechanically gated single-molecule electronic devices.
            
\end{abstract}

\textbf{keywords} Molecular junction, scanning tunnelling microscopy, polythiophene, density functionnal theory

\maketitle

Molecular junctions are perceived as the ultimate step towards the miniaturization of electronic components based on organic materials. While the realisation of integrated 
molecular devices remains a long term goal, understanding the parameters influencing the charge transport through a single molecular bridge is a key step towards complex
functionnal architectures. Experiments based on non-imaging methods have demonstrated that mechanical forces can be used to control the current traversing atomic or molecular junctions \cite{Cuevas1998,Xu2005,Quek2009,Parks2010,Kim2011,Christopher2012,Yoshida2015}. To get further insight into these phenomena a high level of control of the geometrical parameters of the junction is desirable. Thanks to its imaging capability, the scanning tunnelling microscope (STM) was used to probe charge transport through single molecule contacts with extremely high precision \cite{Joachim1995,N'eel2007,Schulze2008,Schulze2008a,Strozecka2009,Berndt2010,Schmaus2011,N'eel2011}. Eventually, this allows determining the influence of atomic scale geometry variations of the molecule--electrode interfaces on the properties of molecular junctions \cite{Temirov2008,Wang2010,Schull2011,Schull2011a,Frederiksen2014}. 
Recently, scanning probe microscopies were also used to probe electronic \cite{Lafferentz2009,Schull2009a,Koch2012}, mechanical \cite{Kawai2014,Wagner2012} and optoelectronic \cite{Reecht2014}
properties of elongated one-dimensional molecular structures, such as molecular dimers \cite{Schull2009a} or single conjugated polymers \cite{Lafferentz2009,Koch2012,Kawai2014,Reecht2014}.
In agreement with earlier observations \cite{William1998,Giese2001,Ho2008,Yamada2008}, these studies report on the exponential decrease of the conductance with the length of the suspended wire.

Here we present a controlled experiment where the conductance of a single molecular wire increases, in some cases by more than one order of magnitude,
despite an increase of its suspended length. To observe this unexpected effect we followed a procedure developped by Lafferentz \textit{et al.} \cite{Lafferentz2009} and used the tip of a STM to progressively lift a polythiophene wire from a Au(111) surface. Conductance traces
recorded during the retraction procedures reveal abrupt increases of the current intensity which we trace back to detachments of the wire sub-units from the surface. Extensive density
functional theory (DFT) simulations that reproduce the overall lifting procedure are performed to interpret the transport data. Based on the agreement between experiment and theory, the sudden increases of conductance are associated to releases of the stress applied on the suspended wire when thiophene units detach from the surface. This
stress relaxation produces a gain in conjugation along the wire and a better electronic coupling with the electrodes, yielding a substantial increase of the transport efficiency of the wire
junction which overcomes the expected loss of conductance due to the wire elongation. This experiment opens the way to electronic devices made of single molecular wires whose transport properties can be tuned mechanically.\\ 

\noindent
The STM experiments were performed on a low temperature ($\approx 4.6$\,K) Omicron apparatus operating under ultrahigh vacuum. The polythiophene wires were synthesized on a
prealably cleaned Au(111) surface, using on-surface polymerisation \cite{Grill2007} of 5,5''-Dibromo-2,2':5',2''-terthiophene following a method described in Ref.\,\citenum{Reecht2013}. 
Etched W tips were annealed and sputtered with Ar$^+$ ions under vacuum. As a final step of preparation, they were gently indented in the sample to cover the apex with gold. Differential 
conductance spectra were recorded with an open feedback loop using lock-in detection with a modulation frequency of 740 Hz and a root-mean-square modulation 
amplitude of 10 to 15 mV. The simulations were performed using an efficient DFT molecular dynamics technique (FIREBALL) \cite{Lewis2011} (see details in Supporting Information\cite{supp}).\\

\begin{figure}[h!]
  \includegraphics[width=10.58cm]{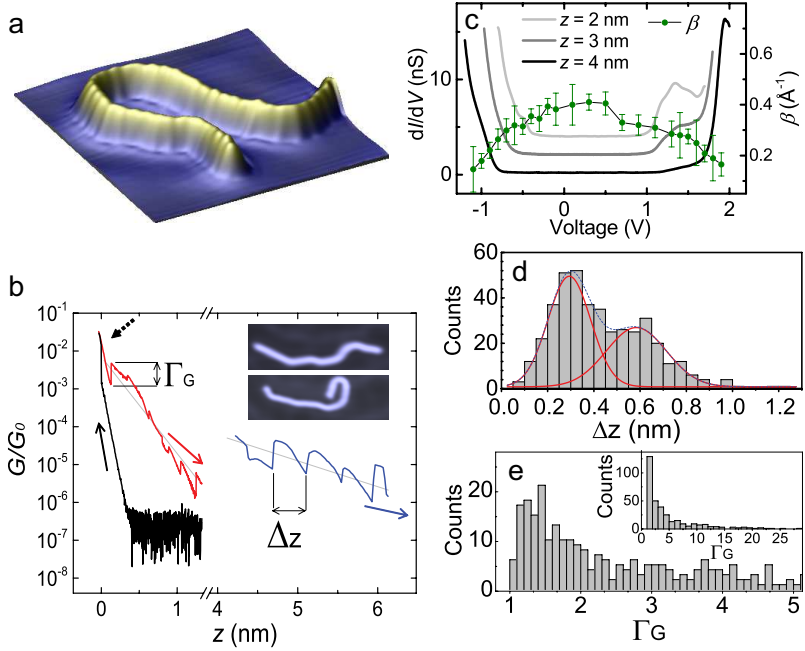}
  \caption{(a) Pseudo-3D STM image of a polythiophene wire on a Au(111) ($7.6 \times 8.3$\,nm$^{2}$, $I=1$\,nA, $V=100$\,mV). (b) Conductance vs tip-sample distance recorded while approaching the metallic tip to the wire extremity (black curve, $V=100$\,mV) and retracting the tip with the wire attached to its apex (red curve, $V=100$\,mV). The blue curve is recorded for a larger tip-sample distance and $V = -1$\,V. The point of contact between the tip and the wire (dashed arrow) defines the origin of the absisca ($z=0$). The STM images in inset ($15.6 \times 5.0$\,nm$^{2}$) show a molecular wire before and after a lifting procedure (lifted from its right extremity). (c) d$I$/d$V$ spectra acquired for different $z$ values and inverse decay length $\beta$ (dots) as a function of $V$ averaged over several wires \cite{note2}. (d) Histogram of the distance separating two successive jumps and (e) of the ratio of the conductances measured after and before the jumps. The histograms in (d) and (e) are constituted,on the basis of 417 and 322 events respectively, and obtained with 50 differents wire junctions.}       
\label{fig1}
\end{figure}

The STM topography in \Figref{fig1}a reveals the intramolecular structure of a characteristic polythiophene wire polymerized on the Au(111) surface. It shows a modulation along the wire with a periodicity of $\approx 0.38$\,nm, which corresponds to the distance separating thiophene units in the polymer \cite{Scifo2006,Reecht2013}. \Figref{fig1}b displays the conductance measured while approaching the STM tip (black curve) to the extremity of the wire in \Figref{fig1}a. For large tip--sample distances ($z=0.5$ to 1\,nm) the conductance is too low to be measured with our experimental setup. Below $z=0.5$\,nm, an exponential decay of the conductance with $z$ is observed as expected for tunnelling transport conditions. The origin of the abscissa ($z= 0$) corresponds to the point of contact between the last atom of the tip and the extremity of the wire, and is characterized by a sudden increase of the conductance (dashed arrow). After reaching this contact configuration, the tip is retracted to its original position. The conductance measured during this procedure (red curve in \Figref{fig1}b) is several orders of magnitude higher than in 
the approach sequence, indicating the presence of the polythiophene wire in the junction. A bias of $V = 0.1$\,V is applied to the junction for the acquisition of these two curves. 
The blue curve corresponds to a conductance trace acquired for a higher voltage ($V = -1$\,V). Both the low and high voltage measurements reveal an overall reduction of the conductance with $z$.
These traces can be fitted as $G(z) \propto G_c exp(-\beta z)$, where $\beta$, the slope of the dashed lines in \Figref{fig1}b, directly reflects the  ability of the wire to transport current {\cite{Ho2008,Yamada2008,Lafferentz2009,Koch2012}. At V = 0.1 V, we measure an average beta of $\approx 0.4 \pm 0.1$ \AA$^{-1}$, in good agreement with calculations \cite{Magoga1997} and close to the value reported for polyfluorene \cite{Lafferentz2009} and graphene nanoribbons \cite{Koch2012}. The slope is milder at elevated voltages, evidencing a more efficient transport of charges. The dependency of $\beta$ with voltage is reproduced in \Figref{fig1}c together with $dI/dV$ spectra acquired for different suspended wire lengths in the junction. As reported recently \cite{Koch2012,Reecht2014}, this graph reveals a correlation between the reduction of $\beta$ and the appearance of a resonance in the $dI/dV$ spectra. In a first approximation, the emerging picture is that the electrons are tunnelling coherently through the wire, as they do through vacuum, but with a strongly enhanced transmission due to the presence of molecular orbitals.

In addition to the overall decrease with $z$, the conductance traces recorded during tip retraction reveal abrupt changes of current. The plot in figure \ref{fig1}d is a histogram of the distance $\Delta z$ between two successive jumps. As highlighted by fits with Gaussian functions, the histogram is dominated by two maxima at $\Delta z = 0.3 \pm 0.1 $\,nm and $\Delta z = 0.6 \pm 0.15$\,nm. 
These values are close to the distances separating a thiophene ring from its nearest neighbor ($0.38$\,nm) or second nearest neighbor ($0.76$\,nm) suggesting that each current jump corresponds to the detachment of one or two thiophene units from the surface \cite{note3}. According to the discussion above and to recent measurements on polyfluorene wires \cite{Lafferentz2009}, the resulting increase of the wire length suspended in the junction should lead to a reduction of the conductance. Surprisingly, the conductance traces in \Figref{fig1}b show the opposite behavior, {\it i.e.}, an important increase of the conductance subsequently to the jump. Figure \ref{fig1}e displays the histogram of the ratio $\Gamma_G$ between the conductance measured after and before the detachment of a wire sub-unit. While the larger occurrence is observed for $1 \leq \Gamma_G \leq 2$, more intense conductance jumps (between 2 to 10) are also frequently observed.

To interpret this unexpected behavior, we performed {\it ab initio} calculations that aim at reproducing the lifting procedure of our experiment. We consider an octothiophene molecule
bridging a gold tip and a gold sample (\Figref{fig2}a). As an initial configuration, one extremity of the wire is anchored {\it via} a covalent C-Au bond to a 35 gold atom STM model tip \cite{note5}. Four thiophene units are suspended into vacuum while the four others are lying flat on the surface.
The junction is relaxed to obtain the equilibrium state and the transmission is calculated using a nonequilibrium Keldysh--Green function formalism which takes multiple scattering into account \cite{Schull2011,Blanco2004,Dappe2014}. In a second step, the tip--sample distance is increased by $0.04$\,nm and the successive relaxation and conductance 
calculations are performed again. This overall process is repeated 13 times, corresponding to a total retraction of $0.52$\,nm of the tip with respect to its initial position (see Video in Supporting Information\cite{supp}).

\begin{figure}[h!]
  \includegraphics[width=10.58cm]{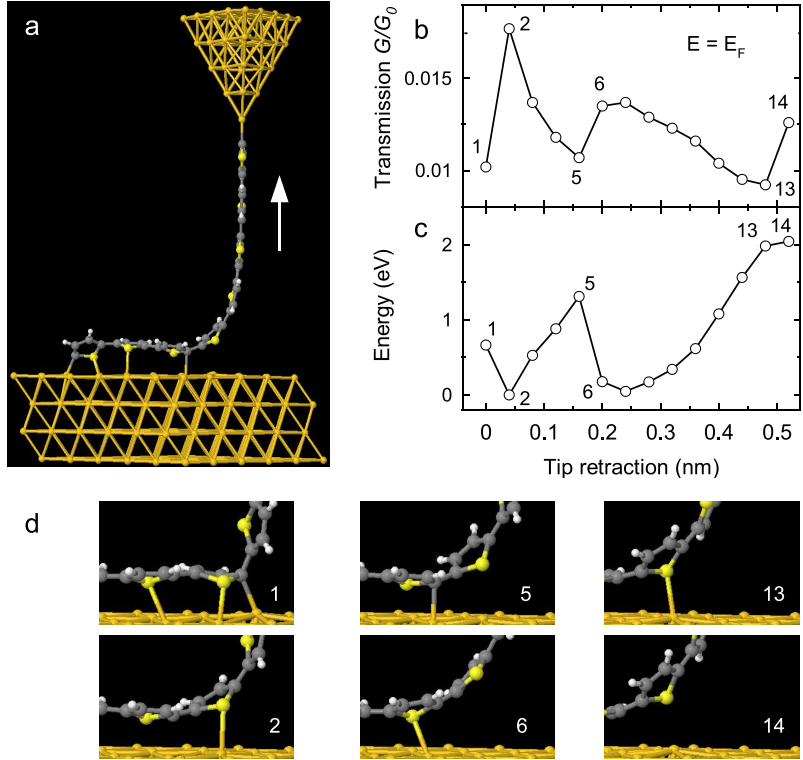}
  \caption{(a) Octothiophene junction considered for the simulations (configuration 5). (b) Transmission at $E=E_F$ and (c) energy calculated for the different configurations (see text for details). (d) Enlarged view of the wire--sample interface for configurations preceding and following a conductance jump in (b).}
\label{fig2}
\end{figure}

Figure \ref{fig2}b and c respectively display the transmission at $E=E_F$ and the energy difference with respect to the most stable calculated configuration as a function of the tip retraction. In three occasions ($1\rightarrow 2$; $5 \rightarrow 6$; $13 \rightarrow 14$), an important increase of the transmission is observed upon configuration changes. 
Both the theoretical values of $\Gamma_G \approx 1.6$ and $\Delta z \approx 0.24$\,nm are close to the 
ones observed in the experiments. 
Interestingly, a decrease of the overall junction energy occurs simultaneously to these conductance jumps, at the exception of the transition between configuration 13 and 14, where
we only observe a stabilization of the energy. Between successive jumps, the transmission (energy) continuously decreases (increases). In \Figref{fig2}d we show 
images of the calculated configurations where we focus on the interface between the wire and the surface before and after a jump. In each case, an important change of the interface geometry is observed, which corresponds either to the cracking of a connexion between a carbon ($1\rightarrow2$; $5\rightarrow6$) or sulfur ($13\rightarrow14$) atom and the Au(111) surface. The emerging picture is that pulling the wire induces a mechanical stress which in turn causes a reduction of the transmission. However, when the stress becomes too large, a part of the wire detaches from the surface ({\it i.e.}, not necessarily a full thiophene unit), resulting in an increase of the conductance. This explains the physical origin of the jumps as well as it identifies a correlation between mechanical stress and conductance. The precise role of the stress on the conductance will be analysed below on the basis of DFT simulations for different configurations.

\begin{figure}[h!]
  \includegraphics[width=8.46 cm]{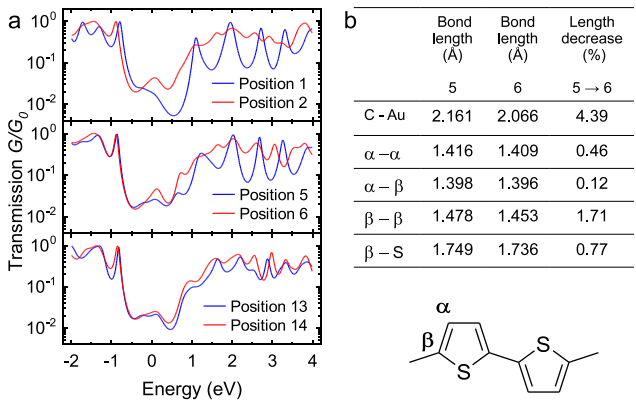}
  \caption{(a) Transmission {\it vs.} energy plots and (b) inter-atomic distances calculated for configurations preceding and following the conductance jumps identified in \Figref{fig2}.}
\label{fig3}
\end{figure}

In figure \ref{fig3}a we compare the energy dependent transmissions calculated for the configurations before and after the three jumps identified in \Figref{fig2}. The highest
occupied molecular orbital (HOMO) (at $E_H \approx -0.8$\,eV) and the lowest unoccupied molecular orbital (LUMO) (at $E_L$ $\approx$ 1.2 eV) can be clearly identified. While the occupied orbitals remain essentially unchanged, an increased broadening of the unoccupied resonances after detachment as well as an increased number of resonances is observed. Both effects may impact the transmission in the gap. These changes suggest that the conjugation along the  suspended part of the wire, the coupling with the tip and the coupling with the part of the wire remaining on the surface, are strongly affected by mechanical stress.

To further explore these aspects we compared the interatomic distance of different bonds before and after a current jump ($5 \rightarrow 6$ case).
After the jump, the length of the Au-C bond between the molecule and the tip is reduced by $\approx 4.4$\,\%, while it is essentially the C--C bonds between neighboring thiophene units ($\beta$--$\beta$ bonds) which are affected within the molecule (bond-length reduction by $\approx 1.7$\,\%). Slightly more difficult to quantify is the effect on the molecule-surface interface where several bonds are involved. To quantify the respective importance of the different bonds, we carried out a more detailed analysis (see Supporting Information \cite{supp}) where the transmission is calculated for junctions where only some sub-parts of the structures are stretched.

These simulations confirm that elongating the molecule--tip connection and the bonds within the wire affect the conductance. However, considering only one of these effects is not sufficient to reproduce the conductance increase after the jump. We conclude that the transmission jumps are due to the detachment of sub-units of the molecular wire from the surface. 
   The mechanical stress release associated to these detachments enhance the conjugation along the wire and reduces the resistance at the molecule--electrode interfaces having for consequence an increased transmission through the wire junction.\\    

The above comparison between theory and experiment gives an explanation for the conductance jumps of moderate intensities ($1 \lesssim  \Gamma_G \lesssim  2$) which correspond to the cases that 
have the largest occurrence in the data. Larger values of $\Gamma_G$, although less frequent, are also observed. Here we speculate that specific geometrical changes upon lifting, such
as twisting of the thiophene units with respect to each other \cite{Venkataraman2006,Dell2013}, may be responsible for the more drastic reduction of the wire conjugation. The sliding of the part
of the wire remaining on the surface occuring during the lifting experiments (\textit{e.g.} inset of \Figref{fig1}b) may also lead to strong reductions of the mechanical stress yielding an improved conductance. These effects are not captured by the present theoretical simulations. This may be due to the specific initial configuration which considers a shorter wire length (for simulation time saving considerations) and an in-line adsorption, which are strong simplifications compared longer wires with more complex adsorption configurations considered experimentally (\textit{e.g.} \Figref{fig1}a).\\

It is also interesting to discuss our results in the scope of similar works with polyfluorene \cite{Lafferentz2009} and graphene nanoribbons \cite{Koch2012} where conductance jumps were not observed. This major difference must be linked to the different chemical nature of the probed polymers. We speculate that the thiophene units are more strongly bound to the gold surface (possibly because of the large chemical affinity between S and Au) than the less reactive sub-structures of the polyfluorene wires and graphene nanoribbons. Consequently, the force required to detach a polythiophene sub-unit should be larger, producing a higher mechanical stress, with the consequence of a stronger reduced conjugation before the detachment. After, the stress is heavily reduced for polythiophene, inducing a large increase of the conductance. This effect is probably much weaker for polyfluorene and graphene nanoribbons.

\begin{figure}[h!]
  \includegraphics[width=8.46cm]{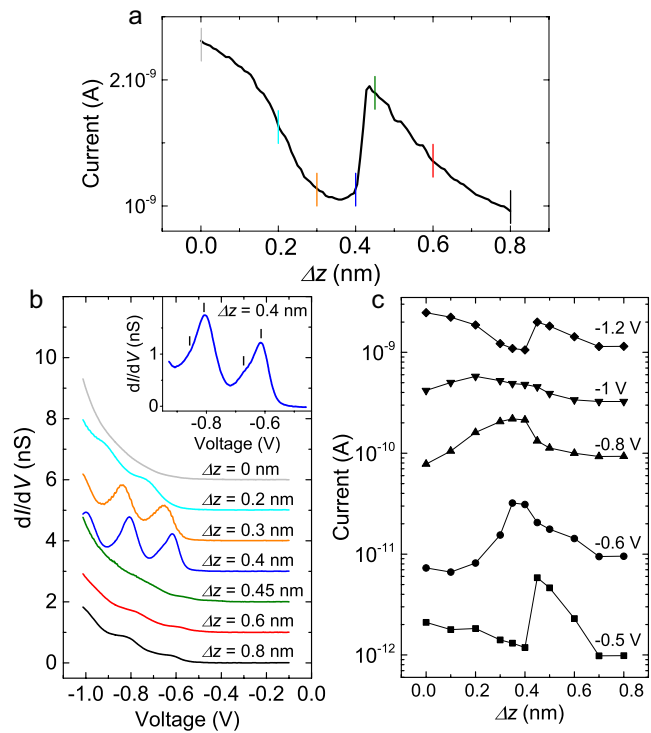}
  \caption{(a) Current ($I$) {\it vs} tip retraction ($\Delta z$) plot showing an experimental conductance jump ($V = 1.2$\,V). (b) $dI/dV$ spectra (shifted vertically for clarity) acquired for different $\Delta z$ values. (c) $I(\Delta z$) plots at different voltages (symbols) reconstructed from the $I(V)$ spectra recorded simultaneously to $dI/dV$ in (b). The connecting lines are guides to the eyes.} 
\label{fig4} 
\end{figure}

In the last part of the manuscript we show how the properties discussed above evolve with voltage. Figure \ref{fig4}a displays a close-up view of a current jump characteristic of the detachment of a wire sub-unit recorded at $V = 1.2$\,V. At this voltage the conductance trace is extremely similar to the one recorded at low voltage (\Figref{fig1}b). In figure \ref{fig4}b we show $dI/dV$ spectra recorded at different positions ($\Delta z$) of this detachment (corresponding to the coloured marks in the spectra \Figref{fig4}a). For $\Delta z= 0$ to $\Delta z = 0.4$\,nm these spectra reveal the progressive apparition of electronic resonances whose widths reduce with increasing mechanical stress on the wire. The spectrum recorded at $\Delta z=0.45$\,nm is the first after the conductance jump, {\it i.e.}, after the stress release. Similarly to the case at $\Delta z=0$, no clear spectral resonances can be distinguished in this spectra. These resonances reappear and get sharper again when the tip is further retracted ($\Delta z = 0.6$ to $0.8$\,nm), a behavior that is similar to the one reported in the zero-volt simulations of \Figref{fig3}. In the inset of figure \ref{fig4}b we can even distinguish vibrational features in the $dI/dV$ spectra \cite{note4} as is generally observed for well decoupled molecules \cite{Ogawa2007,Repp2010,Matino2011}. This is another evidence that stressing the wire reduces the coupling with the electrodes. 
  
These changes in the $dI/dV$ spectral shape also impact $I(z)$ characteristics recorded at voltages close to or at the resonance energies. This is illustrated in \Figref{fig4}c where we represent $I(z)$ curves reconstructed from the $I(V)$ spectra acquired at discrete values of $\Delta z$. For ''out of resonance'' voltage conditions ($V=-1.2$\,V and $V=-0.5$\,V) the sudden conformation change upon detachment yields a current increase. For voltages close to or at resonance with the $dI/dV$ maxima, the current variation upon detachment reveals various behaviors ranging from no current changes ($V = -1$\,V) to current decrease after jump ($-0.6$\,V $\leq V \leq -0.8$\,V). This is the direct consequence of the electronic state narrowing upon stretching \cite{note}.\\  

Our data show that using a STM tip to progressively lift a conjugated polymer wire from a surface provides a unique access to electro-mechanical properties at the single molecular level. Combined with extensive DFT calculations on realistic models, our experiments with polythiophene reveal the response of a molecular junction to mechanical stress. It shows that the stress-induced modifications of both electronic conjugation along the wire and coupling of the wire with the metallic electrodes have a measurable impact on the overall conductance of the wire junction. Finally, these different aspects explain the observation of a transmission increase while the wire suspended in the junction becomes longer. This experiment opens the way to electronic devices made of single molecular wires whose transport properties can be tuned mechanically.

\noindent\textbf{Supporting Information Available}\\
Video 1: Animation showing the calculated junction configurations of a stretched octothiophene wire.\\  
Supporting text 1: Contains detailed information on the simulation methods and Supporting figures (Figure S1 and Figure S2) showing the transmission calculated for junctions where only some sub-parts of the structures are stretched. These materials are available free of charge via the Internet http://pubs.acs.org.

\begin{acknowledgement}
The authors thank J.-G. Faullumel and M. Romeo for technical support. The Agence National de la Recherche (project SMALL'LED No. ANR-14-CE26-0016-01), the Labex NIE (Contract No. ANR-11-LABX-0058\_NIE), the Région Alsace and the International Center for Frontier Research in Chemistry (FRC) are acknowledged for financial support. This work was performed using HPC resources from GENCI-TGCC (Grant no. 2014096813) and GENCI-IDRIS (Grant no. 2014092291).
\end{acknowledgement}

\bibliographystyle{achemso}
\bibliography{mecawire_FS}

\begin{mcitethebibliography}{49}
\bibitem[not()]{note2}
The error bars (y-axis) in \Figref{fig1}c were estimated from the standard
  deviation calculated from all the data available.\relax
\mciteBstWouldAddEndPunctfalse
\mciteSetBstMidEndSepPunct{\mcitedefaultmidpunct}
{}{\mcitedefaultseppunct}\relax
\EndOfBibitem
\bibitem[not()]{note3}
As revealed by STM images in the inset of \Figref{fig1}b, sliding of a part of
  the wire remaining on the surface may occur during the lifting procedure.
  This likely explains the slight mismatch between the experimental and the
  expected inter-ring distances.\relax
\mciteBstWouldAddEndPunctfalse
\mciteSetBstMidEndSepPunct{\mcitedefaultmidpunct}
{}{\mcitedefaultseppunct}\relax
\EndOfBibitem
\bibitem[not()]{note4}
Note that this set of data consitutes an extreme case where the resonances are
  particularly sharp.\relax
\mciteBstWouldAddEndPunctfalse
\mciteSetBstMidEndSepPunct{\mcitedefaultmidpunct}
{}{\mcitedefaultseppunct}\relax
\EndOfBibitem
\bibitem[not()]{note}
The width of the peaks varies from one wire junction to another. Consequently,
  the detailed voltage dependency of the $I$ $vs$ $Z$ plot may differ from
  junction to junction.\relax
\mciteBstWouldAddEndPunctfalse
\mciteSetBstMidEndSepPunct{\mcitedefaultmidpunct}
{}{\mcitedefaultseppunct}\relax
\EndOfBibitem
\end{mcitethebibliography}

\end{document}